\documentclass[journal]{IEEEtran}

\usepackage{cite} 

\usepackage[caption=false,font=footnotesize]{subfig}	
\usepackage{graphicx}

\usepackage{algorithm}
\usepackage{algorithmicx}
\usepackage{algpseudocode}

\usepackage{amsmath,amsfonts,amssymb,amsthm}
\usepackage{mathtools}
\usepackage{mathrsfs}	
\usepackage{subdepth}	

\usepackage{array}	
\usepackage{tabularx}   

\usepackage{hyperref}
\usepackage[sort,compress,capitalise]{cleveref}	
\usepackage{url}

\usepackage[dvipsnames]{xcolor} 
\usepackage{xspace} 

\usepackage{tikz}
\usetikzlibrary{babel}  
\usepackage[straightvoltages,nooldvoltagedirection]{circuitikz}

\usepackage{textcomp}
\usepackage{csquotes}

\usepackage{graphicx}
\usepackage{float}

\usepackage{algorithm}
\usepackage{algorithmicx}
\usepackage{algpseudocode}

\usepackage{amsmath,amsfonts,amssymb,amsthm}
\usepackage{mathtools}
\usepackage{mathrsfs}	
\usepackage{subdepth}	
\usepackage{nicefrac}

\usepackage{array}	
\usepackage{arydshln}   
\usepackage{mathtools}
\usepackage{tabularx}   
\usepackage{tablefootnote}

\usepackage{hyperref}
\usepackage[sort,compress,capitalise]{cleveref}	
\usepackage{url}

\usepackage{xspace} 

\usepackage{tikz}
\usetikzlibrary{fit}
\usetikzlibrary{matrix,decorations.pathreplacing, calc, positioning}
\usetikzlibrary{shapes.symbols}
\usetikzlibrary{babel}  
\usepackage[straightvoltages,nooldvoltagedirection]{circuitikz}
\usetikzlibrary{calc,shapes.geometric}
\usepackage{scalerel}
\usetikzlibrary{arrows.meta}
\usepackage{circuitikz}

\usepackage{nomencl}

{
\theoremstyle{plain}
\newtheorem{Hypothesis}{Hypothesis}

}

{
\theoremstyle{definition}

}

\crefname{Hypothesis}{Hyp.}{Hyps.}
\Crefname{Hypothesis}{Hyp.}{Hyps.}

\crefname{Lemma}{Lemma}{Lemmata}
\Crefname{Lemma}{Lemma}{Lemmata}

\crefname{Definition}{Def.}{Defs.}
\Crefname{Definition}{Def.}{Defs.}

\definecolor{color1}{rgb}{0.510,0.129,0.549}
\definecolor{color2}{rgb}{0.000,0.259,0.549}
\definecolor{color3}{rgb}{0.000,0.451,0.329}
\definecolor{color4}{rgb}{0.961,0.608,0.137}
\definecolor{color5}{rgb}{1.000,0.592,0.592}

\newcommand{\CIGRE}{CIGR{\'E}\xspace}

\newcommand{\DFT}[1][]{DFT\xspace}  
\nomenclature{\DFT}{Discrete Fourier Transform}

\newcommand{\ADC}[1][]{ADC#1\xspace}  
\nomenclature{\ADC}{Analog-to-Digital Converter}
\newcommand{\DAC}[1][]{DAC#1\xspace}  
\nomenclature{\DAC}{Digital-to-Analog Converter}

\newcommand{\LPF}[1][]{LPF#1\xspace}  
\nomenclature{\LPF}{Low-Pass Filter}
\newcommand{\RMS}{RMS\xspace}   
\nomenclature{\RMS}{Root-Mean-Square}

\newcommand{\ctrlPI}{\textup{PI}\xspace}
\newcommand{\ctrlPID}{\textup{PID}\xspace}
\nomenclature{\ctrlPID}{Proportional-Integral-Derivative}
\newcommand{\ctrlPR}{\textup{PR}\xspace}
\nomenclature{\ctrlPR}{Proportional-Resonant}

\newcommand{\ctrlFB}{\textup{\texttt{FB}}\xspace}   
\nomenclature{\ctrlFB}{Feed-Back}
\newcommand{\ctrlFF}{\textup{\texttt{FF}}\xspace}   
\nomenclature{\ctrlFF}{Feed-Forward}
 
\newcommand{\ctrlFT}{\textup{\texttt{FT}}\xspace}   
\nomenclature{\ctrlFT}{Feed-Through}

\newcommand{\MIMO}{MIMO\xspace}	
\nomenclature{\MIMO}{Multi-Input Multi-Output}
\newcommand{\LTI}{LTI\xspace}	
\nomenclature{\LTI}{Linear Time-Invariant}
\newcommand{\LTP}{LTP\xspace}	
\nomenclature{\LTP}{Linear Time-Periodic}
\newcommand{\EHD}{EHD\xspace}	
\nomenclature{\EHD}{Extended Harmonic Domain}
\newcommand{\DPM}{DP\xspace}	
\nomenclature{\DPM}{Dynamic Phasor }
\newcommand{\EMP}{EMP\xspace}	
\nomenclature{\EMP}{Exponentially Modulated Time-Periodic}

\newcommand{\TE}[1][]{\textup{TE#1}\xspace}   
\nomenclature{\TE}{Th{\'e}venin Equivalent}
\newcommand{\NE}[1][]{\textup{NE#1}\xspace}   
\nomenclature{\NE}{Norton Equivalent}

\newcommand{\DAE}[1][]{DAE#1\xspace}		
\nomenclature{\DAE}{Differential-Algebraic Equation}

\newcommand{\MNA}{MNA\xspace}		
\nomenclature{\MNA}{Modified Nodal Analysis}
\newcommand{\MANA}{MANA\xspace}		
\nomenclature{\MANA}{Modified Augmented Nodal Analysis}

\newcommand{\AC}{AC\xspace} 
\newcommand{\DC}{DC\xspace}	

\newcommand{\ADN}[1][]{ADN#1\xspace}	
\nomenclature{\ADN}{Active Distribution Network}
\newcommand{\DER}[1][]{DER#1\xspace}    
\newcommand{\CIDER}[1][]{CI\DER[#1]}    
\nomenclature{\CIDER}{Converter-Interfaced Distributed Energy Resource}
\newcommand{\NIC}[1][]{NIC#1\xspace}         
\nomenclature{\NIC}{Network-Interfacing Converter}

\newcommand{\TDS}{TDS\xspace}	
\nomenclature{\TDS}{Time-Domain Simulation}
\newcommand{\HA}{HA\xspace}		
\nomenclature{\HA}{Harmonic Analysis}
\newcommand{\DHA}{D\HA}			
\nomenclature{\DHA}{Direct Harmonic Analysis}
\newcommand{\IHA}{I\HA}			
\nomenclature{\IHA}{Iterative Harmonic Analysis}
\newcommand{\HDR}[1][]{HDR#1\xspace}	
\nomenclature{\HDR}{Harmonic Domain Response}
\newcommand{\HPF}{HPF\xspace}	
\nomenclature{\HPF}{Harmonic Power Flow}
\newcommand{\HSS}{HSS\xspace}	
\nomenclature{\HSS}{Harmonic State-Space}
\newcommand{\HTF}{HTF\xspace}	
\nomenclature{\HTF}{Harmonic Transfer Function}
\newcommand{\HSA}{HSA\xspace}	
\nomenclature{\HSA}{Harmonic Stability Assessment}

\newcommand{\DI}{DI\xspace}	
\nomenclature{\DI}{Design Invariant}
\newcommand{\CDI}{CDI\xspace}	
\nomenclature{\CDI}{Control-Design Invariant}
\newcommand{\CDV}{CDV\xspace}	
\nomenclature{\CDV}{Control-Design Variant}
\newcommand{\LAP}{LAP\xspace}	
\nomenclature{\LAP}{Linear Assignment Problem}

\newcommand{\EMTP}{EMTP\xspace}		
\nomenclature{\EMTP}{Electromagnetic Transients Program}
\newcommand{\SPICE}{SPICE\xspace}	
\nomenclature{\SPICE}{Simulation Program with Integrated Circuit Emphasis}

\newcommand{\PLL}[1][]{PLL#1\xspace}		
\nomenclature{\PLL}{Phase-Locked Loop}
\newcommand{\PWM}{PWM\xspace}	
\nomenclature{\PWM}{Pulse-Width Modulator}

\newcommand{\KPI}[1][]{KPI#1\xspace}    
\nomenclature{\KPI}{Key Performance Indicator}
\newcommand{\THD}[1][]{THD#1\xspace}    
\nomenclature{\THD}{Total Harmonic Distortion}
\newcommand{\DFS}[1][]{DFS#1\xspace}	
\nomenclature{\DFS}{Double Fourier Series}

\newcommand{\Abs}[1]{\left|#1\right|}

\newcommand{\Exp}[1]{\exp\left(#1\right)}

\newcommand{\diag}{\operatorname{diag}}

\newcommand{\grd}{\gamma}
\newcommand{\pwr}{\pi}
\newcommand{\ctrl}{\kappa}
\newcommand{\act}{\alpha}

\newcommand{\trafo}{\tau}
\newcommand{\refr}{\rho}
\newcommand{\spt}{\sigma}
\newcommand{\opt}{o}

\newcommand{\Y}{\mathbf{Y}}     

\newcommand{\rsc}{q}

\newcommand{\phsA}{\textup{\texttt{A}}}

\newcommand{\phsABC}{\textup{\texttt{ABC}}\xspace}

\newcommand{\cmpD}{\textup{\texttt{D}}}
\newcommand{\cmpQ}{\textup{\texttt{Q}}}
\newcommand{\cmpDQ}{\textup{\texttt{DQ}}}
\newcommand{\cmpZ}{\textup{\texttt{Z}}}
\newcommand{\cmpDQZ}{\textup{\texttt{DQZ}}\xspace}

\newcommand{\seqP}{\textup{\texttt{P}}}
\newcommand{\seqN}{\textup{\texttt{N}}}
\newcommand{\seqH}{\textup{\texttt{H}}}

\newcommand{\IT}{\mathbf{i}}

\begin{document}

\title{\huge{%
    Harmonic Stability Analysis of Microgrids with Converter-Interfaced Distributed Energy Resources,\\
	Part II: Case Studies
}}

\author{%
	Johanna~Kristin~Maria~Becker,~\IEEEmembership{Student Member,~IEEE},
	Andreas~Martin~Kettner,~\IEEEmembership{Member,~IEEE},\\
	and~Mario~Paolone,~\IEEEmembership{Fellow,~IEEE}%
	\thanks{J. Becker and M. Paolone are with the Distributed Electrical Systems Laboratory at the {\'E}cole Polytechnique F{\'e}d{\'e}rale de Lausanne (EPFL) in CH-1015 Lausanne, Switzerland (E-mail: \{johanna.becker, mario.paolone\}@epfl.ch).}%
	\thanks{A. Kettner is with PSI NEPLAN AG, 8700 Küsnacht, Switzerland (E-mail: andreas.kettner@neplan.ch).}%
	\thanks{This work was funded by the Schweizerischer Nationalfonds (SNF, Swiss National Science Foundation) via the National Research Programme NRP~70 ``Energy Turnaround'' (projects nr. 173661 and 197060).
    }%
}

\maketitle







\begin{abstract}
    In Part~I of this paper a method for the \emph{Harmonic Stability Assessment} (\HSA) of power systems with a high share of \emph{Converter-Interfaced Distributed Energy Resources} (\CIDER[s]) was proposed.
    Specifically, the \emph{Harmonic State-Space} (\HSS) model of a generic power system is derived through combination of the components' \HSS models.
    The \HSS models of \CIDER[s] and grid are based on \emph{Linear Time-Periodic} (\LTP) models, capable of representing the coupling between different harmonics.
	In Part~II, the \HSA of a grid-forming, and two grid-following \CIDER[s] (i.e., ex- and including the \DC-side modelling) is performed.
    More precisely, the classification of the eigenvalues, the impact of the maximum harmonic order on the locations of the eigenvalues, and the sensitivity curves of the eigenvalues w.r.t.~to control parameters are provided.
    These analyses allow to study the physical meaning and origin of the \CIDER[s]' eigenvalues.
	Additionally, the \HSA is performed for a representative example system derived from the \CIGRE low-voltage benchmark system.
    A case of harmonic instability is identified through the system eigenvalues, and validated with \emph{Time-Domain Simulations}~(\TDS) in Simulink.
    It is demonstrated that, as opposed to stability analyses based on \emph{Linear Time-Invariant} (\LTI) models, the \HSA is suitable for the detection of harmonic instability.
\end{abstract}





\section{Introduction}
\label{sec:intro}

%
%
%



\IEEEPARstart{T}{he} method for the \emph{Harmonic Stability Assessment}~(\HSA), as introduced in Part~I of this paper, is applied to (i) individual resources, and (ii)~entire power systems.
Specifically, the \HSA is performed for \emph{Converter-Interfaced Distributed Energy Resources} (\CIDER[s]), as well as a small yet realistic example system, which is derived from the \CIGRE low-voltage benchmark system~\cite{Rep:2014:CIGRE}.
All examples are framed in the context of distribution systems and, hence, the assumptions stated in Section~III-B do apply.

Recall from Part~I that the \HSA method investigates the eigenvalues of the \emph{Harmonic State-Space}~(\HSS) models.
The analyses performed for individual \CIDER[s] are three-fold.
First, the impact of the maximum harmonic order on the locations of the eigenvalues is illustrated.
Second, the classification of the eigenvalues into \emph{Control-Design Invariant}~(\CDI), \emph{Control-Design Variant}~(\CDV) and \emph{Design Invariant}~(\DI) is studied in detail.
Additionally, the sensitivity curves of the eigenvalues w.r.t.~to specific control parameters are given.
Thus, it is demonstrated that the proposed tools and analyses allow to (i) quantify the differences between \emph{Linear Time-Periodic} (\LTP) and \emph{Linear Time-Invariant} (\LTI) models of \CIDER[s], (ii) identify the physical meaning of the \CIDER's eigenvalues through modal analysis, providing an explanation for the so-called spurious eigenvalues, and (iii) examine stability characteristics of the \CIDER w.r.t. their tuning.

The \HSA of a representative example system, which is derived from the \CIGRE low-voltage benchmark system, is performed.
More precisely, the eigenvalues of the closed-loop system are analysed in detail, and a comparison w.r.t. the open-loop system components is performed.
A case of harmonic instability is identified through the system eigenvalues and validated with \emph{Time-Domain Simulations} (\TDS) in Simulink.
It is also shown that this instability cannot be observed with the conventional stability criterion (i.e., the eigenvalues of the corresponding \LTI system).
The studies performed in this paper show that the modelling framework for harmonic analysis proposed in \cite{jrn:2020:kettner-becker:HPF-1,jrn:2020:kettner-becker:HPF-2} and extended in \cite{jrn:2022:becker}, is suitable for \HSA of generic \CIDER models, as well as entire power systems.



The remainder of Part~II is organised as follows:
In \cref{sec:hsa-rsc}, the \HSS models of the \CIDER[s], that were introduced in \cite{jrn:2020:kettner-becker:HPF-2} and \cite{jrn:2022:becker},
are analysed in detail in the context of \HSA.
\cref{sec:hsa-sys} gives the \HSA for a small but representative example system (i.e., with parameters based on the \CIGRE low-voltage benchmark microgrid).
Finally, the conclusions are drawn in \cref{sec:conclusion}.

\section{Harmonic Stability Assessment of the CIDERs}
\label{sec:hsa-rsc}


This section gives an overview on the \HSA of the individual \CIDER models.
First, the modelling and parameters of the adopted \CIDER[s] are recalled in \cref{sec:hsa-rsc:model-param}.
Based on these models, several studies are conducted:
(i) the eigenvalues are classified w.r.t. their sensitivity to model parameters in \cref{sec:hsa-rsc:classification},
(ii) the impact of the maximum harmonic order on the eigenvalues of the \CIDER[s] is analysed in \cref{sec:hsa-rsc:truncation},
and (iii) the eigenvalue loci w.r.t.~to the controller parameters (i.e., feedback gains) are assessed in \cref{sec:hsa-rsc:sensitivity}.

For the sake of conciseness, each study is presented based on the \CIDER model that is most suited for the phenomenon being discussed.
Studies and discussions covering all three \CIDER models can be found in \cite{ths:2024:becker}.

\subsection{Resource Modelling and Parameters}
\label{sec:hsa-rsc:model-param}

The resource models introduced in \cite{jrn:2020:kettner-becker:HPF-2} and \cite{jrn:2022:becker} are employed for the \HSA in this paper.
Specifically, three types of \CIDER models are considered: (i) the grid-forming and, (ii) the grid-following \CIDER that considers only \AC-side characteristics (introduced in Section~III-B and III-C of \cite{jrn:2020:kettner-becker:HPF-2}, respectively), and (iii) the grid-following \CIDER that includes the modelling of the \DC side (proposed in Section~IV of \cite{jrn:2022:becker}).
\cref{fig:CIDER:model} shows a detailed schema of a generic \CIDER.
The modelling of the internal response (i.e., power hardware, control software and coordinate transformations) stays unchanged as compared to \cite{jrn:2020:kettner-becker:HPF-2} and \cite{jrn:2022:becker}.
The difference, as opposed to the \CIDER models previously introduced, is the representation of the reference calculation as a small-signal model.
This applies to both of the grid-following \CIDER[s], namely the reference calculation being the PQ law and is detailed in Appendix~\ref{app:cider:lib-rsc}.

\begin{figure}[t]
    \centering
    {

\tikzstyle{block-big}=[rectangle,draw=black,minimum width=1.4cm,minimum height=1.0cm,inner sep=0pt]
\tikzstyle{conversion}=[rectangle,draw=black,minimum width=0.9cm,minimum height=0.6cm,inner sep=0pt]
\tikzstyle{transformation}=[rectangle,draw=black,minimum width=0.7cm,minimum height=0.7cm,inner sep=0pt]
\tikzstyle{controller}=[rectangle,draw=black,minimum width=1.0cm,minimum height=1.0cm,inner sep=0pt]

\tikzstyle{dot}=[circle,draw=black,fill=black,inner sep=1pt]
\tikzstyle{signal}=[-latex]

\definecolor{myRed}{rgb}{1 0 0}
\definecolor{myGreen}{rgb}{0 0 1}

\scriptsize

\begin{tikzpicture}
    \def\dx{1.0}
    \def\dy{1.0}

    \draw[dashed,draw=myRed] (3.8*\dx,1.5*\dy) to (3.8*\dx,-4.5*\dy);
	\draw[signal,draw=myRed] (3.6*\dx,1.3*\dy) to (3.3*\dx,1.3*\dy);
	\node[text=myRed] at (2.3*\dx,1.3*\dy)
	{%
	    \begin{tabular}{c}
	        Internal Response
	    \end{tabular}
	};
    
    
    \node[block-big] (A) at (0,0)
    {%
    \begin{tabular}{c}
        Actuator\\
        $\act$
    \end{tabular}
    };
    
    \node[block-big] (F) at ($(A)+(2.2*\dx,0)$)
    {%
    \begin{tabular}{c}
        Filter\\
        Stages
    \end{tabular}
    };
    \node[dot] (FO) at ($(F)-(0,1.25*\dy)$) {};
    \draw[signal] (A.east) to (F.west);
    
    \node[dot] (PCI) at ($(F)+(2.5*\dx,0)$) {};
    \coordinate (PCO) at ($(PCI)-(0,1.25*\dy)$);
    
    \node[transformation] (TGI) at ($(PCI)+(\dx,0)$) {$\tau_{\pwr|\gamma}$};
    \node[transformation] (TGO) at ($(PCO)+(\dx,0)$) {$\tau_{\gamma|\pwr}$};
    
    \node (GI) at ($(TGI)+(1.2*\dx,0)$) {$\mathbf{w}_{\grd}$};
    \node (GO) at ($(TGO)+(1.2*\dx,0)$) {$\mathbf{y}_{\grd}$};
    \node (G) at ($0.5*(GI)+0.5*(GO)$) {Grid $\grd$};
    
    \draw[signal] (GI.west) to (TGI.east);
    \draw[signal] (TGO.east) to (GO.west);
    
    \draw (TGI.west) to (PCI.east);
    \draw[signal] (PCI.east) to node[near end,above]{$\mathbf{w}_{\pwr}$} (F.east);
    \draw[signal] (FO.east) to  (TGO.west); 
    
    
    
    
    
    
    
    \node[transformation] (TBW) at ($(A)-(0,2.0*\dy)$) {$\tau_{\pwr|\ctrl}$};
    \draw[signal] (TBW.north) to node[midway,right]{$\mathbf{u}_{\pwr}$} (A.south);

    \node[transformation] (TFWI) at ($(F)-(0,2.0*\dy)$) {$\trafo_{\ctrl|\pwr}$};
    \draw[signal] (F.south) to node[midway,right]{$\mathbf{y}_{\pwr}$} (TFWI.north);
    \node[transformation] (TFWO) at ($(PCI)-(0,2.0*\dy)$) {$\tau_{\ctrl|\pwr}$};
    \draw[signal] (PCI.south) to (TFWO.north);
    
    
    
    
    \node[block-big] (K) at ($(TFWI)-1.5*(0,\dy)$)
    {%
    \begin{tabular}{c}
        Controller\\
        Stages
    \end{tabular}
    };
    \node[block-big] (R) at ($(TFWO)-1.5*(0,\dy)$)
    {%
    \begin{tabular}{c}
        Reference\\
        $\refr$
    \end{tabular}
    };
    
    \node (SP) at ($(R)+(2.2*\dx,0)$) {$\mathbf{w}_{\spt}$};
    \node at ($(SP)+(0,0.4*\dy)$) {Setpoint $\spt$};
    
    \draw[signal] (TFWI.south) to node[midway,right]{$\mathbf{u}_{\ctrl}$} (K.north);
    \draw[signal] (TFWO.south) to node[midway,right]{$\mathbf{w}_{\refr}$} (R.north);
    
    \draw[signal] (SP.west) to (R.east);
    \draw[signal] (R.west) to node[midway,above]{$\mathbf{w}_{\ctrl}$} (K.east);
    \draw[signal] (K.west) to node[midway,above]{$\mathbf{y}_{\ctrl}$} ($(TBW)-1.5*(0,\dy)$) to (TBW.south);
    
    
    
    
    
    \draw[dashdotted] ($(F.north east)+0.2*(\dx,\dy)$)
        to ($(F.south east)+0.2*(\dx,-\dy)$)
        to ($(A.south west)+0.2*(-\dx,-\dy)$)
        to ($(A.north west)+0.2*(-\dx,\dy)$)
        to ($(F.north east)+0.2*(\dx,\dy)$);
    \node at ($0.5*(F.north)+0.5*(A.north)+0.4*(0,\dy)$) {Power Hardware $\pwr$};

    \draw[dashdotted] ($(K.north east)+0.2*(\dx,\dy)$)
        to ($(K.south east)+0.2*(\dx,-\dy)$)
        to ($(K.south west)+2.2*(-\dx,0)+0.2*(-\dx,-\dy)$)
        to ($(K.north west)+2.2*(-\dx,0)+0.2*(-\dx,\dy)$)
        to ($(K.north east)+0.2*(\dx,\dy)$);
    \node at ($(K.south)+1.1*(-\dx,0)+0.4*(0,-\dy)$) {Control Software $\ctrl$};
    
    
    
    
    
    
    
    
    
    
    
\end{tikzpicture}

}
    \caption
    {%
        Block diagram of a generic \CIDER.
        The internal response is consists of the power hardware $\pwr$, control software $\ctrl$ and coordinate transformations $\tau_{\cdot|\cdot}$.
        Together with the reference calculation $\refr$ and additional transformations (e.g., accounting for changes in circuit configurations) it builds the \CIDER response at the point of connection to the grid $\grd$.
    }
    \label{fig:CIDER:model}
\end{figure}

The exemplary parameters for the grid-forming and grid-following \CIDER[s] are listed in \cref{tab:cider-forming:parameters,tab:cider-following:parameters,tab:cider-following-dc:parameters}, respectively.
The setpoints are $V_\spt=230\,\text{V-\RMS}$ and $f_\spt=50\,\text{Hz}$ for the grid-forming \CIDER, and $P_\spt=-50\,\text{kW}$ and $Q_\spt=-16.4\,\text{kVAr}$  for the grid-following one.
For the grid-following \CIDER including the \DC-side characteristics, the same $PQ$ setpoints are chosen, and additionally the \DC voltage reference is set to $V_{\delta}^*=900\,\text{V}$.
The control parameters of the \CIDER[s] are tuned by investigating the eigenvalue loci obtained through the \HSA as will be introduced in \cref{sec:hsa-rsc:sensitivity}%
\footnote{%
	In practice, the \ctrlPI controllers of \CIDER[s] are tuned employing techniques like the symmetrical and/or magnitude optimum \cite{Jrn:Umland:1990,Bk:Teodorescu:2011}, or pole placement~\cite{Jrn:Liserre:2005}.
}.
In this particular case, the goal of the tuning was not to achieve optimal performance but to facilitate a meaningful case study.

\begin{table}[t]
	\centering
	\caption
	{%
		Parameters of the Grid-Forming Resource (Rated Power $40\,\text{kVA}$)
	}
	\label{tab:cider-forming:parameters}
	{
	
	\renewcommand{\arraystretch}{1.1}
	\setlength{\tabcolsep}{0.15cm}
	
	\begin{tabular}{lccccc}
		\hline
		Filter stage
		&	$L$/$C$
		&	$R$/$G$
		&	$K_{\ctrlFB}$
		&	$T_{\ctrlFB}$
		&	$K_{\ctrlFT}$
		\\
		\hline
		Inductor ($\alpha$)
		&	0.49\,mH
		&	1.53\,m$\Omega$
		&	4
		&	8E-4
		&	1
		\\
		Capacitor ($\varphi$)
		&	60.2\,$\mu$F
		&	0\,S
		&	1.5
		&	1E-3
		&	0
		\\
		\hline
	\end{tabular}
	
}	
\end{table}

\begin{table}[t]
	\centering
	\caption
	{%
		Parameters of the Grid-Following Resource (Rated Power $60\,\text{kVA}$)
	}
	\label{tab:cider-following:parameters}
	{
	
	\renewcommand{\arraystretch}{1.1}
	\setlength{\tabcolsep}{0.15cm}
	
	\begin{tabular}{lccccc}
		\hline
		Filter stage
		&	$L$/$C$
		&	$R$/$G$
		&	$K_{\ctrlFB}$
		&	$T_{\ctrlFB}$
		&	$K_{\ctrlFT}$
		\\
		\hline
		Actuator-side inductor ($\alpha$)
		&	325~$\mu$H
		&	1.02~m$\Omega$
		&	5
		&	5E-4
		&	1
		\\
		Capacitor ($\varphi$)
		&	90.3~$\mu$F
		&	0~S
		&	1
		&	8E-4
		&	0
		\\
		Grid-side inductor ($\gamma$)
		&	325~$\mu$H
		&	1.02~m$\Omega$
		&	1
		&	1E-3
		&	1
		\\
		\hline
	\end{tabular}
	
}
\end{table}

\begin{table}[t]
	\centering
	\caption
	{%
		Parameters of the Grid-Following Resource Including the \DC~Side (Rated Power $60\,\text{kVA}$)
	}
	\label{tab:cider-following-dc:parameters}
	{

\renewcommand{\arraystretch}{1.1}
\setlength{\tabcolsep}{0.15cm}

\begin{tabular}{lccccc}
	\hline
		Filter stage
	&	$L$/$C$
	&	$R$/$G$
	&	$K_{\ctrlFB}$
	&	$T_{\ctrlFB}$
	&	$K_{\ctrlFT}$
	\\
	\hline
	    DC-Link Capacitor ($\delta$)
	&	310~$\mu$F
	&	0~S
	&	10
	&	1E-2
	&	0
	\\
		Actuator-side inductor ($\alpha$)
	&	325~$\mu$H
	&	1.02~m$\Omega$
	&	2.5
	&	5E-4
	&	1
	\\
	    Capacitor ($\varphi$)
	&	90.3~$\mu$F
	&	0~S
	&	0.8
	&	1E-3
	&	0
	\\
		Grid-side inductor ($\gamma$)
	&	325~$\mu$H
	&	1.02~m$\Omega$
	&	0.5
	&	5E-3
	&	1
	\\
	\hline
\end{tabular}

}
\end{table}

\begin{figure}[t]
	\centering
	{
	
	\scriptsize
	\ctikzset{bipoles/length=0.8cm}
	
	\begin{circuitikz}[european]
		
		\def\x{0.9}
		\def\y{0.7}
		
		\coordinate (SN) at (0,0);
		\coordinate (SP) at ($(SN)+(0,2*\y)$);
		\coordinate (ON) at ($(SN)+(2*\x,0)$);
		\coordinate (OP) at ($(SP)+(2*\x,0)$);
		\coordinate (I) at ($0.5*(SN)+0.5*(SP)-(\x,0)$);
		\coordinate (R) at ($(I)+(5*\x,0)$);
		
		
		
		\node[rectangle,draw=black,minimum size=2.0cm] (Res) at (R)
		{%
			\begin{tabular}{c}
				\CIDER\\
				(detailed model)
			\end{tabular}
		};
		\draw ($(Res.west)+(0,\y)$) to[short] (OP);
		\draw ($(Res.west)+(0,-\y)$) to[short] (ON);
		
		
		\draw (ON)
		to[short] (SN)
		to[voltage source = $V_{\TE}$] (SP)
		to[R = $Z_{\TE}$] (OP);
		\draw (ON) to[open,o-o,v=$ $] (OP);
		
	\end{circuitikz}
	
}
	\caption
	{%
		Test setup for the validation of the individual \CIDER models.
		The resource is represented by a detailed state-space model, and the power system by a \TE (see \cref{tab:TE:parameters,tab:TE:harmonics}).
	}
	\label{fig:val-rsc:setup}
\end{figure}
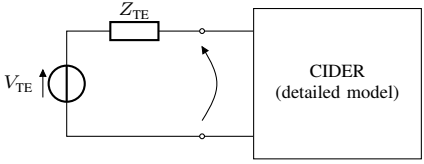

Employing the \HSS model of the \CIDER[s], the eigenvalues of system matrix $\mathbf{\tilde{A}}_{\rsc}$ as introduced in (22)--(23) in Section~III-D of Part~I are evaluated.
Note that the system matrix includes only the characteristics of the internal response of the \CIDER (cf. \cref{fig:CIDER:model}), i.e., the matrices of the reference calculation do not enter the system matrix of the individual \CIDER[s].
Additionally, as was shown in Section~II-C of \cite{jrn:2022:becker}, the matrices of the individual \CIDER model may be a function of the operating point $\mathbf{\hat{Y}}_{\opt}$, due to the linearisations performed during the derivation of the \LTP model of the \CIDER.
For the \HSA of the individual \CIDER, if the derivation of the \HSS model requires an operating point, it is taken from spectra obtained from \TDS%
\footnote{%
    Alternatively, one can run a preliminary \HPF study and use the obtained spectra for the operating point.
    In fact, this approach will be used shortly for the \HSA of the entire system.
}.
For the \TDS, the test setup shown in \cref{fig:val-rsc:setup} is used.
It consists of two parts: a \emph{Th{\'e}venin Equivalent} (\TE) that represents the grid, and a detailed model of the \CIDER under investigation.
The \TE impedance is characterized by typical short-circuit parameters of a power distribution grid, which are given in \cref{tab:TE:parameters}.
The \TE voltage source injects harmonics, whose levels are given in \cref{tab:TE:harmonics}.
These levels are set according to the limits specified in the standard BS-EN-50160:2000 \cite{Std:BSI-EN-50160:2000}. %
In line with this standard, harmonics up to order 25 (i.e., 1.25~kHz) are considered in the analysis.

\begin{table}[t]
	\centering
	\caption{Short-Circuit Parameters of the Th{\'e}venin Equivalent}
	\label{tab:TE:parameters}
	{

\renewcommand{\arraystretch}{1.1}

\begin{tabular}{cccl}
    \hline
        Parameter
    &   Resource
    &   System
    &   Description
    \\
    &   Analysis
    &   Analysis
    \\
    \hline
        $V_{n}$
    &   230\,V-\RMS
    &   230\,V-\RMS
    &   Nominal voltage 
    \\
        $S_{\mathit{sc}}$
    &   267\,kW
    &   3.85\,MW
    &   Short-circuit power
    \\
        $\Abs{Z_{\mathit{sc}}}$
    &   195\,m$\Omega$
    &   13.7\,m$\Omega$
    &   Short-circuit impedance
    \\
        $R_{\mathit{sc}}/X_{\mathit{sc}}$
    &   6.207
    &   0.271
    &   Resistance-to-reactance ratio
    \\
    \hline
\end{tabular}
}
\end{table}

\begin{table}[t]
	\centering
	\caption{Harmonic Voltages of the Th{\'e}venin Equivalent (see \cite{Std:BSI-EN-50160:2000}).}
	\label{tab:TE:harmonics}
	{
	\renewcommand{\arraystretch}{1.1}
	\begin{tabular}{ccr}
		\hline
		$h$
		&   $|V_{\TE,h}|$
		&   \multicolumn{1}{c}{$\angle V_{\TE,h}$}
		\\
		\hline
		1
		&   1.0\,p.u.
		&   0\,rad
		\\
		5
		&   6.0\,\%
		&   $\pi$/8\,rad
		\\
		7
		&   5.0\,\%
		&   $\pi$/12\,rad
		\\ 
		11
		&   3.5\,\%
		&   $\pi$/16\,rad
		\\
		13
		&   3.0\,\%
		&   $\pi$/8\,rad
		\\
		17
		&   2.0\%
		&   $\pi$/12\,rad
		\\
		19
		&   1.5\,\%
		&   $\pi$/16\,rad
		\\
		23
		&   1.5\,\%
		&   $\pi$/16\,rad
		\\
		\hline
	\end{tabular}
}

\end{table}


\subsection{Impact of the Maximum Harmonic Order}
\label{sec:hsa-rsc:truncation}

\subsubsection{Methodology}

The impact of the maximum harmonic order on the location of the eigenvalues is assessed based on the grid-following \CIDER that includes the \DC-side modelling.
To this end, the eigenvalues of the system matrix are computed for different values of $h_{max}$.

Additionally, the eigenvalues of the \LTP models are compared to the ones of the corresponding \LTI model.
In general, when modelling \CIDER[s] with \LTI models, all parts of the \CIDER need to be represented in the same frame of reference (e.g, \cmpDQ~components). 
By consequence, the transformation matrices between the power hardware and control software simplify to identity. 
As to the \LTP models, the operating point for deriving the \LTI models is taken from the spectra of \TDS.

In order to quantify by how much the eigenvalues move for different maximum harmonic orders, the following metric is introduced.
Let $\lambda_{i,\LTI}$ be an eigenvalue of the \LTI system, with $i$ ranging from 1 to $n_{\LTI}$ and $n_{\LTI}$ being the size of the \LTI system matrix.
Next, for a given $h_{max}$, find the $n_{\LTI}$ eigenvalues of the \LTP system that are located closest to those of the \LTI system%
\footnote{%
    Notably, in case two modes of eigenvalues are located close to each other, there is no general guarantee that this procedure will find the correct matching.
}.
Denote these $n_{\LTI}$ eigenvalues of the \LTP model by $\lambda_{i,\LTP}^{h_{max}}$, where $i = 1\ldots n_{\LTI}$.
Note that $\lambda_{i,\LTP}^{h_{max}}$ is a subset of all eigenvalues of the \LTP model, since the size of the system matrix satisfies $n_{\LTP}>n_{\LTI}$ in general.
The distance between each corresponding pair of \LTI and \LTP eigenvalues is calculated as:
\begin{align}
	d_{i}(h_{max}) = ||\lambda_{i,\LTI}-\lambda_{i,\LTP}^{h_{max}}||_1
\end{align}
The metric for similarity between the \LTI and \LTP systems for a given $h_{max}$ is the maximum of these distances:
\begin{align}
	d(h_{max}) = \max_{i}(d_{i}(h_{max}))
	\label{eq:hsa:similarityMetric}
\end{align}
Notably, $d(h_{max})$ is (i) a measure of similarity between the \LTP and \LTI model and (ii) an indicator for the maximum harmonic order required to represent all relevant phenomena of the system.
With regard to the latter, one needs to chose $h_{max}$ large enough so that $d(h_{max})$ reaches a stable value (i.e., $d$ could not be reasonably decreased by further increasing $h_{max}$).

\subsubsection{Results and Discussion}

\begin{figure}[t]
	\centering
    \includegraphics[width=0.9\linewidth]{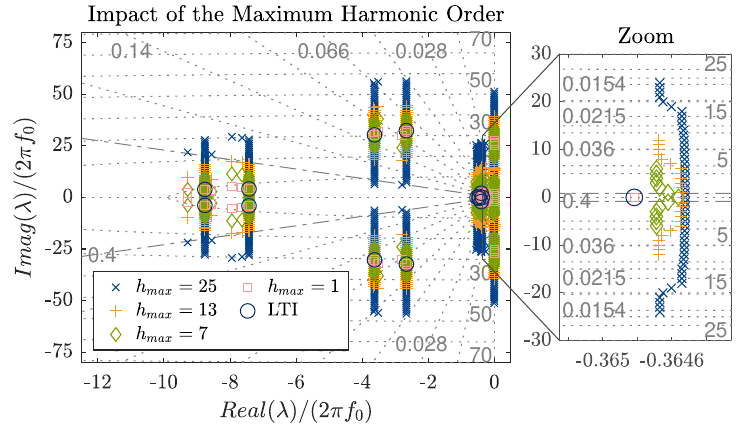}
	
	\caption
	{%
		Analysis of the impact of the maximum harmonic order on the location of the \LTP eigenvalues for the grid-following \CIDER that models the \DC side. 
	}
	\label{fig:hsa-rsc:truncation}
\end{figure}

\cref{fig:hsa-rsc:truncation} shows the eigenvalues of the grid-following \CIDER with \DC-side modelling for maximum harmonic orders equal to $1, 7, 13$ and $25$, as well as for the \LTI model.
One can clearly see that the sets of eigenvalues are clustered in lines (i.e., with similar real part), one of which is displayed in detail in the zoomed-in excerpts of \cref{fig:hsa-rsc:truncation}.
More precisely, each mode in time-domain corresponds to one cluster of eigenvalues in the harmonic domain.
When increasing the maximum harmonic order, the number of eigenvalues in such a cluster increases too.
This is what one would expect, corresponding to the increase in size of the system matrix.
For the grid-following \CIDER that models the \DC side, the location of the eigenvalues changes when increasing the maximum harmonic order (cf. \cref{fig:hsa-rsc:truncation}).

\begin{figure}[t]
	\centering
	\includegraphics[width=0.9\linewidth]{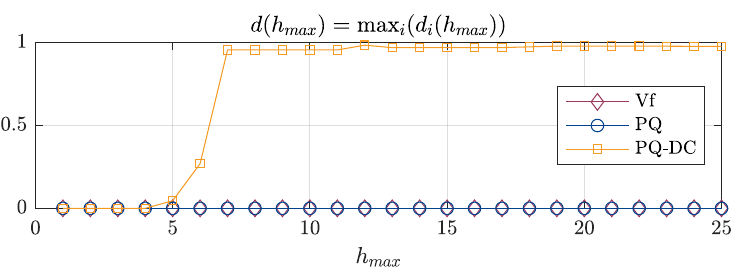}	
	\caption
	{%
		Analysis of the impact of $h_{max}$ on the location of the \LTP compared to the \LTI eigenvalues for the three \CIDER[s].
		The \CIDER[s] being analysed are the grid-forming and grid-following \CIDER[s] that only model the \AC-side characteristics, and the grid-following \CIDER that includes the \DC-side modelling.
	}
	\label{fig:hsa-rsc:ltp-vs-lti-abs}
\end{figure}

The observation made above is confirmed by \cref{fig:hsa-rsc:ltp-vs-lti-abs}, which shows the maximum distance between the \LTI and \LTP eigenvalues $d(h_{max})$ as defined in \eqref{eq:hsa:similarityMetric} for all three \CIDER models.
For the two \CIDER[s] which only model the \AC-side characteristics $d(h_{max})$ is constant at zero.
Hence, increasing the maximum harmonic order does not affect the location of the eigenvalues and their real part matches with the one of the \LTI eigenvalues.
This indicates that, in this particular case, there are no time-varying phenomena present in the \LTP model, which would be missed by the \LTI model.

For the grid-following \CIDER which includes the \DC-side characteristics, $d(h_{max})$ in \cref{fig:hsa-rsc:ltp-vs-lti-abs} increases with $h_{max}$.
For a $h_{max}$ up to four, the difference between the \LTI and the \LTP eigenvalues is close to zero.
When increasing $h_{max}$ further the difference increases.
The changes in \cref{fig:hsa-rsc:ltp-vs-lti-abs} can be explained by the entries of the operating point.
In fact, the grid-following \CIDER that models the \DC side includes a linearization of the internal response w.r.t.~a time-periodic operating point.
When increasing $h_{max}$, more and more of the time-periodic entries are considered in the derivation of the system matrix.
Hence, whenever a new nonzero entry of the operating point is considered, the location of the eigenvalues changes.
For the case of the grid-following \CIDER with \DC side, the biggest changes of $d(h_{max})$ occur around a maximum harmonic order of $h_{max} = 5-7$.
This indicates that the nonlinear \CIDER behaviour strongly impacts this part of the harmonic spectrum.
This phenomenon can be explained by the high levels of those harmonics (cf. \cref{tab:TE:harmonics}).
This analysis clearly demonstrates the difference in how \CIDER[s] are represented by \LTI models compared to \LTP models.


\subsection{Classification of the \LTP eigenvalues}
\label{sec:hsa-rsc:classification}

\subsubsection{Methodology}
As discussed in Section~VI-C of Part~I, eigenvalues of a resource comprising a plant and its controller can be classified based on their sensitivity to model parameters into \CDI, \CDV, and \DI eigenvalues.
To categorise these, eigenvalues are computed for two sets of parameters.
Specifically, in order to identify \DI eigenvalues, all parameters of the \CIDER are varied, and unchanged eigenvalues are classified as \DI.
For \CDI and \CDV eigenvalues, control software parameters of a \CIDER are altered, with unchanged eigenvalues labelled as \CDI and those that shift as \CDV.
The sets of eigenvalues calculated for each variation are ordered using the \emph{Linear Assignment Problem} (\LAP) (cf. Section~VI-B in Part~I). 

\subsubsection{Results and Discussion}

\begin{figure*}[t]
	\centering
	\subfloat[]
	{%
		\centering
		\includegraphics[width=0.5\linewidth]{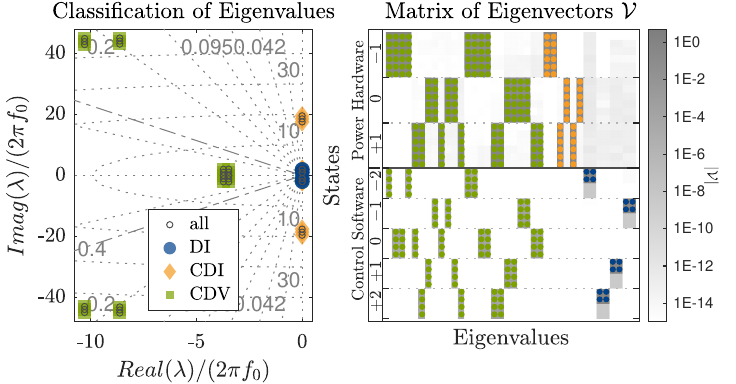}
		\label{fig:hsa-rsc:classification:frm}
	}
	\subfloat[]
	{%
		\centering
		{
	\tikzstyle{square}=[rectangle,draw=black,minimum height=0.3cm,minimum width=0.3cm,inner sep=0mm]
	\tikzstyle{UG1}=[]
	\tikzstyle{UG3}=[dashed]
	\tikzstyle{signalP}=[-latex,color3,line width=0.5pt]
	\tikzstyle{signalN}=[-latex,color1,line width=0.5pt]
	\tikzstyle{signalH}=[-latex,color4,line width=0.5pt]
	
	\begin{circuitikz}
		\scriptsize
		
		\def\BlockSize{1.0}	
		\def\X{1.25}	
		\def\Y{2}	
		\def\dlX{0.3}
		\def\dlY{0.3}
		\def\dload{0.2}
		
		
		\node (T1) at ($(\X/2,5*\dlY)$) {$\phsABC$ to $\cmpDQZ$ Transformation:};
		
		\node (P1) at ($(0,2.1*\dlY)$) {$h_{\phsABC}$};
		\node (K1) at ($(\X,3.1*\dlY)$) {$h_{\cmpDQZ}$};
		
				
		\node[square] (PHhm1_o) at (0,\dlY) {-1};
		\node[square] (PHh0) at (0,0) {\phantom{-}0};
		\node[square] (PHhp1_o) at (0,-\dlY) {\phantom{-}1};
		
		
		\node[square] (CShm2_o) at (\X,2*\dlY) {-2};
		\node[square] (CShm1) at (\X,\dlY) {-1};
		\node[square] (CSh0) at (\X,0) {\phantom{-}0};
		\node[square] (CShp1) at (\X,-\dlY) {\phantom{-}1};
		\node[square] (CShp2_o) at (\X,-2*\dlY) {\phantom{-}2};
		
		\draw[signalP] (PHhm1_o.east) to (CShm2_o.west);
		\draw[signalP] (PHh0.east) to (CShm1.west);
		\draw[signalP] (PHhp1_o.east) to (CSh0.west);

		\draw[signalN] (PHhm1_o.east) to (CSh0.west);
		\draw[signalN] (PHh0.east) to (CShp1.west);
		\draw[signalN] (PHhp1_o.east) to (CShp2_o.west);
		
		\draw[signalH] (PHhm1_o.east) to (CShm1.west);
		\draw[signalH] (PHh0.east) to (CSh0.west);
		\draw[signalH] (PHhp1_o.east) to (CShp1.west);

		
		\coordinate (O) at ($(2.75*\X,0)$);
		
		\node (T2) at ($(O)+(\X/2,5.0*\dlY)$) {$\cmpDQZ$ to $\phsABC$ Transformation:};
		
		\node (P2) at ($(O)+(\X,4.1*\dlY)$) {$h_{\phsABC}$};
		\node (K2) at ($(O)+(0,3.1*\dlY)$) {$h_{\cmpDQZ}$};
		
		
		\node[square,draw = gray] (PHhm3) at ($(O)+(\X,3*\dlY)$) {\textcolor{gray}{-3}};
		\node[square,draw = gray] (PHhm2) at ($(O)+(\X,2*\dlY)$) {\textcolor{gray}{-2}};
		\node[square] (PHhm1) at ($(O)+(\X,\dlY)$) {-1};
		\node[square] (PHh0) at ($(O)+(\X,0)$) {\phantom{-}0};
		\node[square] (PHhp1) at ($(O)+(\X,-\dlY)$) {\phantom{-}1};
		\node[square,draw = gray] (PHhp2) at ($(O)+(\X,-2*\dlY)$) {\textcolor{gray}{\phantom{-}2}};
		\node[square,draw = gray] (PHhp3) at ($(O)+(\X,-3*\dlY)$) {\textcolor{gray}{\phantom{-}3}};
		
		
		\node[square] (CShm2) at ($(O)+(0,2*\dlY)$) {-2};
		\node[square] (CShm1) at ($(O)+(0,\dlY)$) {-1};
		\node[square] (CSh0) at ($(O)+(0,0)$) {\phantom{-}0};
		\node[square] (CShp1) at ($(O)+(0,-\dlY)$) {\phantom{-}1};
		\node[square] (CShp2) at ($(O)+(0,-2*\dlY)$) {\phantom{-}2};
		
		\draw[signalP] (CShm2.east) to (PHhm1.west);
		\draw[signalP] (CShm1.east) to (PHh0.west);
		\draw[signalP] (CSh0.east) to (PHhp1.west);
		\draw[signalP,dashed] (CShp1.east) to (PHhp2.west);
		\draw[signalP,dashed] (CShp2.east) to (PHhp3.west);
		
		\draw[signalN,dashed] (CShm2.east) to (PHhm3.west);
		\draw[signalN,dashed] (CShm1.east) to (PHhm2.west);
		\draw[signalN] (CSh0.east) to (PHhm1.west);
		\draw[signalN] (CShp1.east) to (PHh0.west);
		\draw[signalN] (CShp2.east) to (PHhp1.west);
		
		\draw[signalH,dashed] (CShm2.east) to (PHhm2.west);
		\draw[signalH] (CShm1.east) to (PHhm1.west);
		\draw[signalH] (CSh0.east) to (PHh0.west);
		\draw[signalH] (CShp1.east) to (PHhp1.west);
		\draw[signalH,dashed] (CShp2.east) to (PHhp2.west);
		
		
		\coordinate (Leg) at (-0.2*\X,-4.5*\dlY);
		\matrix [draw=white,below right,fill=white] at (Leg) {
			\node [signalP,label=right:~~~~~~Pos. Seq.] {}; &
			\node [signalN,label=right:~~~~~~Neg. Seq.] {}; &
			\node [signalH,label=right:~~~~~~Hom. Seq.] {}; \\
		};
		
		\def\dl{1.47*\X}
		\def\do{-0.85*\dlY}
		
		\draw[signalP] ($(Leg)+(0.7*\dlX,\do)$) to node[]{} ($(Leg)+(2.5*\dlX,\do)$);
		\draw[signalN] ($(Leg)+(0.7*\dlX,\do)+(\dl,0)$) to node[]{} ($(Leg)+(2.5*\dlX,\do)+(\dl,0)$);
		\draw[signalH] ($(Leg)+(0.7*\dlX,\do)+(2*\dl,0)$) to node[]{} ($(Leg)+(2.5*\dlX,\do)+(2*\dl,0)$);
		
		
		\draw[dotted,lightgray] ($(PHhm1_o.north west)-\dlX*(1,0)$) to ($(PHhm1.north east)+\dlX*(2.5,0)$);
		\draw[dotted,lightgray] ($(PHhp1_o.south west)-\dlX*(1,0)$) to ($(PHhp1.south east)+\dlX*(2.5,0)$);
		
		\node at ($(PHhm1.north east)+\dlX*(2.5,0.4)$) {$-h_{max,\phsABC}$};
		\node at ($(PHhp1.south east)+\dlX*(2.5,-0.5)$) {$h_{max,\phsABC}$};
		
		\draw[dashdotted,lightgray] ($(CShm2_o.north west)-\dlX*(6,0)$) to ($(CShm2.north east)+\dlX*(8,0)$);
		\draw[dashdotted,lightgray] ($(CShp2_o.south west)-\dlX*(6,0)$) to ($(CShp2.south east)+\dlX*(8,0)$);
		
		\node at ($(CShm2.north east)+\dlX*(8,0.4)$) {$-h_{max,\cmpDQZ}$};
		\node at ($(CShp2.south east)+\dlX*(8,-0.5)$) {$h_{max,\cmpDQZ}$};

	\end{circuitikz}
}
		\label{fig:hsa:schema:transformation}
	}
 
	\subfloat[]
	{%
		\centering
		\includegraphics[width=0.7\linewidth]{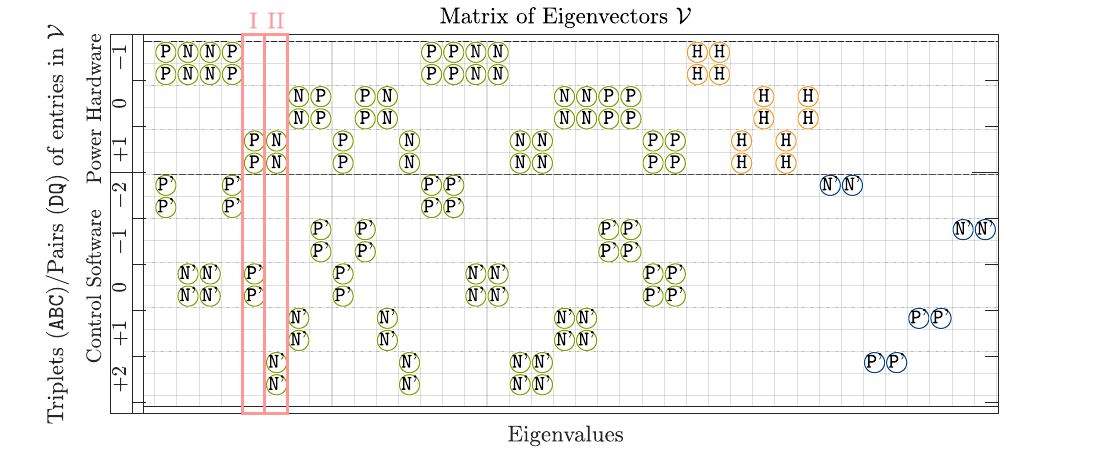}
		\label{fig:hsa:frm:classification:Vdetail}
	}

	\caption
	{%
		Classification of the eigenvalues of the grid-forming \CIDER.
		\cref{fig:hsa-rsc:classification:frm}: Eigenvalues on the left-hand side and eigenvector matrix $\boldsymbol{\mathcal{V}}$ on the right-hand side.
        The dark grey line in the plot of the eigenvector matrix indicates the separation between the states of the power hardware and control software.
        The individual states are further partitioned w.r.t. their harmonic order by the dotted light grey lines.      
		\cref{fig:hsa:schema:transformation}: Mapping of the harmonic sequences for a transformation from \phsABC coordinates to \cmpDQZ components and vice versa.
		\cref{fig:hsa:frm:classification:Vdetail}: Representation of the entries of $\boldsymbol{\mathcal{V}}$ by triplets (\phsABC) and pairs (\cmpDQ) in the sequence domain.
	}
	\label{fig:hsa-rsc:classification}
\end{figure*}

\cref{fig:hsa-rsc:classification} shows the classification of the \LTP eigenvalues for the grid-forming \CIDER.
For the sake of illustration, the analysis is performed for $h_{max}=1$%
\footnote{
    Notably, even though only frequencies up to the fundamental frequency are considered, the \HSS model still differs from a corresponding \LTI model in how the frequency coupling is represented.
}.
While the generalisation of the methodology to higher values of $h_{max}$ is straightforward (i.e., the patterns of eigenvectors w.r.t.~the eigenvalues are analogous), the visibility of the plots given the available space deteriorates.

The subfigures in \cref{fig:hsa-rsc:classification:frm} display the \DI, \CDI, and \CDV eigenvalues on the left, and the magnitude of the eigenvector matrix $\boldsymbol{\mathcal{V}}$ entries in logarithmic scale on the right.
Entries exceeding a threshold of $1$E-8~p.u. in magnitude are highlighted in the colour corresponding to the eigenvalue classification.
The rows (x-axis) and columns (y-axis) of the eigenvector matrix represent the states and eigenvalues of the \CIDER model, respectively.
As detailed in \cite{jrn:2020:kettner-becker:HPF-1}, a \CIDER's harmonic-domain state vector is composed of the power hardware and the control software states.
Hence, the y-axis is partitioned accordingly.
It is further partitioned w.r.t.~to the harmonic orders of the states.
Note that it was chosen to represent the control software of the \CIDER[s] by $h_{max}+1$ harmonics.

In order to understand the physical meaning of the eigenvalues, an in-depth examination of the eigenvectors is performed.
To this end, the entries of the eigenvectors in \cref{fig:hsa:frm:classification:Vdetail} are grouped by triplets (i.e., \phsABC coordinates) for the part associated with the power hardware and pairs (i.e., \cmpDQ~components) for the part associated with the control software.
The \phsABC triplets are classified in the sequence domain as positive (\seqP), negative (\seqN) and homopolar (\seqH) sequences.
Similarly, the \cmpDQ pairs are interpreted as an equivalent positive sequence (\seqP') when the \cmpQ~component lags behind the \cmpD~component by ninety degrees, and an equivalent negative sequence (\seqN') when the \cmpQ~component leads the \cmpD~component by ninety degrees \cite{Jrn:Rygg:2017}.

For ease of understanding, the mapping of harmonic sequence components for a transformation from phase coordinates (i.e., \phsABC frame) in the power hardware to direct/quadrature components (i.e., \cmpDQ(\cmpZ) frame) in the control software is illustrated in \cref{fig:hsa:schema:transformation}.
More precisely, the harmonic order of positive and negative sequences is decremented and incremented by 1, respectively, whereas the harmonic order of homopolar sequences remains unchanged~\cite{Jrn:Rygg:2017}.
Note the different maximum harmonic orders considered for the two coordinate frames (i.e., $h_{max,\phsABC}=1$ for the \phsABC coordinates and $h_{max,\cmpDQZ}=h_{max,\phsABC}+1=2$ for the \cmpDQZ components).
In the following, the detailed analysis of the eigenvalues is first discussed for the \CDV, then for the \CDI and finally for the \DI eigenvalues.

\paragraph{\CDV Eigenvalues}
\cref{fig:hsa-rsc:classification:frm} shows four sets of \CDV eigenvalues.
Each of these sets is composed of three complex conjugate eigenvalues, i.e., one per harmonic order.
The corresponding eigenvectors in the right-hand subplot have nonzero entries (in green) for both parts of the state vector (i.e., these eigenvalues influence states both in the power hardware and control software).
By inspecting the highlighted area marked as I in \cref{fig:hsa:frm:classification:Vdetail}, one observes that the entries associated with the power hardware correspond to \phsABC-triplets of positive sequences located at $h = 1$.
Similarly, for the control software, the entries at harmonic order $h = 0$ correspond to equivalent positive sequences.
In the highlighted area denoted with II, the entries of the eigenvector associated with the power hardware correspond to \phsABC triplets of negative sequences located at $h=1$.
Additionally, for the control software, the entries at harmonic order $h = 2$ correspond to equivalent negative sequences.
These observations fit with the theoretical mapping of harmonic sequence components for a transformation from \phsABC coordinates to \cmpDQ(\cmpZ) components in \cref{fig:hsa:schema:transformation}.
The harmonic order is decremented for positive-sequence components and incremented for negative-sequence components.

\paragraph{\CDI Eigenvalues}
Only one set of \CDI eigenvalues occurs in the eigenvalue graph in \cref{fig:hsa-rsc:classification:frm}.
The corresponding eigenvectors have nonzero entries (in yellow) only in the entries associated with the power hardware.
This is due to the \CDI eigenvalues being invariant w.r.t.~changes in the control parameters (i.e., by definition).
A deeper understanding of this fact can be obtained by inspection of \cref{fig:hsa:frm:classification:Vdetail}.
The nonzero entries of the eigenvectors associated with the \CDI eigenvalues exhibit values corresponding to a homopolar sequence.
Recall that the power hardware is modelled in \phsABC coordinates and the control software in \cmpDQ~components (i.e., the \cmpZ~component is not considered in the controller).
By consequence, any homopolar sequences originating from the power hardware are not mapped to the control software.

\paragraph{\DI Eigenvalues}
Finally, one pair of \DI eigenvalues with real part equal to zero is observed in \cref{fig:hsa-rsc:classification:frm}.
Their eigenvectors are solely nonzero (in blue) in the entries associated with the control software.
In particular, these entries correspond to the innermost controller stage (i.e., the one providing the voltage reference to the actuator of the power hardware).
As illustrated on the right-hand side of \cref{fig:hsa:schema:transformation}, positive sequences at $h = -h_{max}$ and $h = -(h_{max}+1)$, as well as negative sequences at $h = h_{max}$ and $h = h_{max}+1$ are cut off during the \cmpDQ(\cmpZ)-to-\phsABC transformation.
By consequence, such eigenvalues can neither be properly controlled nor analysed if they arise.

In line with the remarks regarding spurious eigenvalues in Section~VI-C in Part~I of this paper, these observations substantiate that the \DI eigenvalues are artefacts of the model rather than genuine eigenvalues of the system.
Therefore, although the \DI eigenvalues possess real parts equal to zero, they do not indicate a problem of system stability.
That is, although they are borderline (un)stable and invariant w.r.t.~parameter changes, their origin is clear and their influence is marginal.
Naturally, this reasoning is only valid if the maximum harmonic order $h_{max}$ chosen for the \HSA is high enough to cover all relevant phenomena, which was discussed in the previous section.
The topic of the choice of $h_{max}$ and its influence on the resource eigenvalues was discussed in \cref{sec:hsa-rsc:truncation}.


\subsection{Sensitivity Analysis of the Eigenvalues}
\label{sec:hsa-rsc:sensitivity}

\subsubsection{Methodology}

In order to determine the sensitivity of the eigenvalues w.r.t.~specific controller parameters, it suffices to vary only these parameters individually (i.e., as opposed to the previous analysis where several or all parameters are varied simultaneously).
Then, the eigenvalues of the original and modified system are sorted employing the \LAP, following the same principles as described in Section~VI-B of Part~I.
This procedure is repeated for several incremental changes of the controller parameters under investigation.
In this way, one can trace the eigenvalue loci in the Laplace plane.

This approach is illustrated through the case of the grid-following \CIDER which only models the \AC-side characteristics.
To this end, the sensitivity of the eigenvalues w.r.t.~the change of the controller feedback gains $K_{\ctrlFB,\act}$ of the \CIDER are analysed (i.e., the feedback gain of the inner-most controller stage).
The controller gain is gradually incremented by $1$\% of its initial value (as given in \cref{tab:cider-following:parameters}) for $N = 70$ iterations.

The sensitivities of the eigenvalues w.r.t.~control parameters are useful in the context of parameter-tuning problems.
More precisely, one can select the optimal locations of the eigenvalues (i.e., w.r.t.~a given objective function) while ensuring that the resource is stable.
In this paper, it is shown how to obtain the sensitivities and perform a preliminary analysis of the resource and system behaviour w.r.t.~to changes of selected control parameters.
The optimal tuning of the parameters remains for future work.

\subsubsection{Results and Discussion}

\begin{figure}[t]
	\centering
		\includegraphics[width=0.9\linewidth]{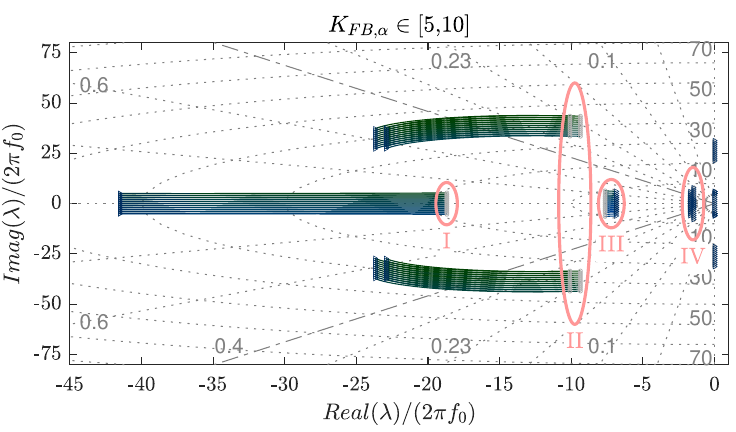}
	
	\caption
	{%
		Sensitivity curves w.r.t.~the feedback gain $K_{\ctrlFB}$ of the controller stage associated with 
        the actuator-side inductance 
        of the grid-following \CIDER, which models only the \AC-side characteristics.
		The maximum harmonic order considered for this analysis is $h_{max}=5$.
	}
	\label{fig:hsa-rsc:sensitivity:flw}
\end{figure}

\cref{fig:hsa-rsc:sensitivity:flw} shows the eigenvalue loci of the grid-following \CIDER model which considers only the \AC-side characteristics.
To simplify the discussion, the eigenvalues are ordered into areas I to IV.
The tuning of $K_{\ctrlFB,\act}$ of the inner-most controller stage leads to a very low sensitivity of the eigenvalues in area I in \cref{fig:hsa-rsc:sensitivity:flw} (i.e., they do almost not change).
The eigenvalues in area II-IV on the other hand, are significantly moved in direction of the negative real axis.
Hence, in order to achieve better tuning of the \CIDER, it is beneficial to increase $K_{\ctrlFB,\act}$ of the inner-most controller stage.
By increasing the gain up to $10$, the eigenvalues of area II can be moved into the area of damping factors bigger than $0.4$.

\section{Harmonic Stability Assessment of Entire Power Systems}
\label{sec:hsa-sys}

This section discusses the \HSA of an entire microgrid.
In \cref{sec:hsa-sys:setup}, the test system is introduced and the steady-state operating point is characterized by means of a \HPF study.
A comparison  of the system eigenvalues to the eigenvalues of the open-loop resources, as well as to the \LTI eigenvalues is performed in \cref{sec:hsa-sys:eig-det}.
Finally, in \cref{sec:hsa-sys:instab} a case of harmonic instability is identified using the loci of the system eigenvalues and validated through \TDS with Simulink.
For the sake of conciseness, the classification of the system eigenvalues similar to \cref{sec:hsa-rsc:classification} is not discussed in this paper.
A detailed analysis in this respect can be found in \cite{ths:2024:becker}.


\subsection{Test System Setup}
\label{sec:hsa-sys:setup}

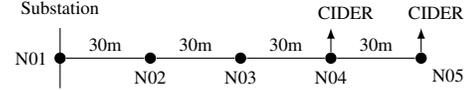
\begin{figure}[t]
	\centering
	{
	\tikzstyle{bus}=[circle,fill=black,minimum size=1.5mm,inner sep=0mm]
	\tikzstyle{UG1}=[]
	\tikzstyle{UG3}=[dashed]
	\tikzstyle{load}=[-latex]
	
	\begin{circuitikz}
		\scriptsize
		
		\def\BlockSize{1.0}	
		\def\X{5}	
		\def\Y{5}	
		\def\dlX{1.2}
		\def\dlY{1.3}
		\def\dload{0.4}
		
		
		\node[bus,label={left:N01}] (N1) at (0,0) {};
		\node[bus,label={below:N02}] (N2) at (\dlX,0) {};
		\node[bus,label={below:N03}] (N3) at (2*\dlX,0) {};
		\node[bus,label={below:N04}] (N4) at (3*\dlX,0) {};
		\node[bus,label={below right:N05}] (N5) at (4*\dlX,0) {};
		
		
		\draw[UG1] (N1) to node[midway,above]{30m} (N2) {};
		\draw[UG1] (N2) to node[midway,above]{30m} (N3) {};
		\draw[UG1] (N3) to node[midway,above]{30m} (N4) {};
		\draw[UG1] (N4) to node[midway,above]{30m} (N5) {};
		
		\draw[load] (N4) to 
		($(N4)+\dload*(0,1)$);
		\node (L1) at ($(N4)+\dload*(0.5,1.5)$) {\CIDER};
		\draw[load] (N5) to 
		($(N5)+\dload*(0,1)$);
		\node (L2) at ($(N5)+\dload*(0.5,1.5)$) {\CIDER};
		
		\draw[-] ($(N1)+\dload*(0,-1)$) to ($(N1)+\dload*(0,1)$);
		\node[label={above:Substation}] (Substation) at ($(N1)+\dload*(0,1)$) {};
		
%
		
	\end{circuitikz}
}
	\caption
	{%
		Schematic diagram of the test system used for the \HSA.
		Its line lengths and cable parameters are based on the \CIGRE low-voltage benchmark microgrid \cite{Rep:2014:CIGRE}.
        Cable parameters are given in \cref{tab:grid:parameters}.
	}
	\label{fig:5bus-grid:schematic}
\end{figure}

The proposed \HSA is applied to the test system shown in \cref{fig:5bus-grid:schematic}.
For ease of understanding (i.e., a manageable number of eigenvalues), a small-scale example system is chosen.
The parameters are taken from the \CIGRE low-voltage benchmark microgrid~\cite{Rep:2014:CIGRE}.
The \HSS model is derived for the complete system model, and used to assess the system's eigenvalues.

The test system is characterized as follows.
The substation is located in node N01.
Its short-circuit parameters, are listed in \cref{tab:TE:parameters}.
Recall from \cref{sec:hsa-rsc} that, the \TE voltage source injects harmonics, whose levels are given in \cref{tab:TE:harmonics} and set according to the limits specified in the standard BS-EN-50160:2000 \cite{Std:BSI-EN-50160:2000}.
All lines are built from underground cables,
whose sequence parameters are given in \cref{tab:grid:parameters}.
Two grid-following \CIDER[s], of which only the \AC-side characteristics are modelled, are connected at nodes N04 and N05.
Their parameters are the same as for the resource validation, see \cref{tab:cider-following:parameters}.
In line with the analysis in \cref{sec:hsa-rsc:sensitivity}, the parameters of one of the \CIDER[s] are retuned in order to achieve better overall damping.
More precisely, the feedback gain of the inner-most controller stage $K_{\ctrlFB,\act}$ for the \CIDER at N04 is increased to $16$.
The setpoints for both grid-following \CIDER[s] are set to $P_\spt=-50\,\text{kW}$ and $Q_\spt=-16.4\,\text{kVAr}$.

\begin{table}[t]
	\centering
	\caption{Sequence Parameters of the Lines in the Test System.}
	\label{tab:grid:parameters}
	{
	\renewcommand{\arraystretch}{1.1}
	\setlength{\tabcolsep}{0.15cm}
	
	\begin{tabular}{cccccc}
		\hline
        $R_{+}/R_{-}$ 
		&   $R_{0}$ 
		&   $L_{+}/L_{-}$ 
		&   $L_{0}$ 
		&   $C_{+}/C_{-}$ 
		&   $C_{0}$ 
		\\
		\hline
        0.162~$\Omega$
		&   0.529~$\Omega$
		&   0.262~mH
		&   1.185~mH
		&   637~nF
		&   388~nF
		\\
		\hline
	\end{tabular}
}
\end{table}

\begin{figure}[t]
	\centering
	\includegraphics[width=0.9\linewidth]{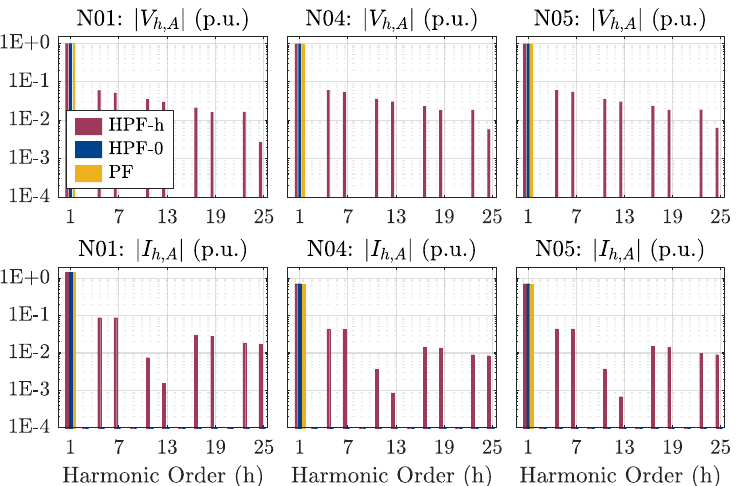}
	\caption{
		Results of the \HPF for a system with harmonic distortion (\HPF-h), the \HPF for a system with zero harmonic distortion (\HPF-0) and a conventional power-flow at the fundamental frequency (PF).
        The magnitudes of nodal voltages and currents of phase $\phsA$ at the nodes where resources are connected are depicted on the top and bottom, respectively.
	}
	\label{fig:5bus:nodal145}
\end{figure}

In order to derive the \HSS model of the system, the harmonic operating point of the system is derived through the \HPF introduced in \cite{jrn:2020:kettner-becker:HPF-1}.
Notably, the operating points needed for the system under consideration are the harmonic phasors of the nodal voltages at the nodes where the grid-following \CIDER[s] are connected.
As explained in detail in Appendix~\ref{app:cider:lib-rsc}, the nodal voltages are used in the linearisation of the reference calculation of the grid-following \CIDER[s].
Throughout this section, the eigenvalues obtained from the \HSS model are compared w.r.t. the ones obtained from a classical \LTI model.
Analogously to the \LTP model, the \LTI model requires the nodal voltages at nodes, where the grid-following \CIDER[s] are connected, for the linearisation of the reference calculation.
In order to derive the nodal voltages for the \LTI case, a conventional power-flow is conducted at the fundamental frequency.
The operating points which have been obtained for the described cases are displayed in \cref{fig:5bus:nodal145}.
Is it worth noting that the fundamental component of the nodal voltages is close to $1$~p.u. in magnitude, which means that the system is lightly loaded.

\subsection{Detailed Analysis of the System Eigenvalues}
\label{sec:hsa-sys:eig-det}

\subsubsection{Methodology}

For the detailed analysis of the system eigenvalues, a maximum harmonic order of $h_{max} = 25$ is considered.
The aim of this analysis is to give a general understanding of the different groups of eigenvalues being analysed.
Due to the large amount of eigenvalues, as well as their wide distribution, a separation into two sets of eigenvalues is introduced, i.e., one each associated with the grid and the resources, respectively.

The set of system eigenvalues associated with the grid is briefly discussed by comparing them to the eigenvalues of the open-loop grid model.
Subsequently, a more detailed discussion of the set of eigenvalues associated with the resources is given.
First, they are compared to the eigenvalues of the \TE and the two \CIDER[s] (i.e., the open-loop resource models).
Second, the impact of the harmonics on the system eigenvalues is analysed.
To this end, the eigenvalues of a system that includes harmonics as in \cref{tab:TE:harmonics} is compared to one without harmonic distortion.
Notably, for the two analyses different operating points (i.e., computed through the \HPF with and without harmonic distortion) need to be considered when deriving the \HSS model of the system.
Third, the two sets of system eigenvalues are compared to the ones obtained by an \LTI system model.
Notably, \LTI models are entirely represented in \cmpDQ~components, in order to circumvent the consideration of the time-periodic transformation matrices.
Thus, the difference between \LTP and \LTI is the modelling of the power hardware in \phsABC coordinates and \cmpDQ~components, respectively, as well as the specific consideration of the transformation blocks in the \LTP model.

The comparison of the eigenvalues obtained from the system with harmonic distortion, the system without harmonic distortion, and the \LTI model, allows to understand (i) whether an instability occurs due to harmonic distortion, and (ii) whether it can be detected by the conventional stability criteria (i.e., by the eigenvalues of the \LTI system).

\subsubsection{Results and Discussion}

\begin{figure}[t]
	\centering
	
		\includegraphics[width=0.9\linewidth]{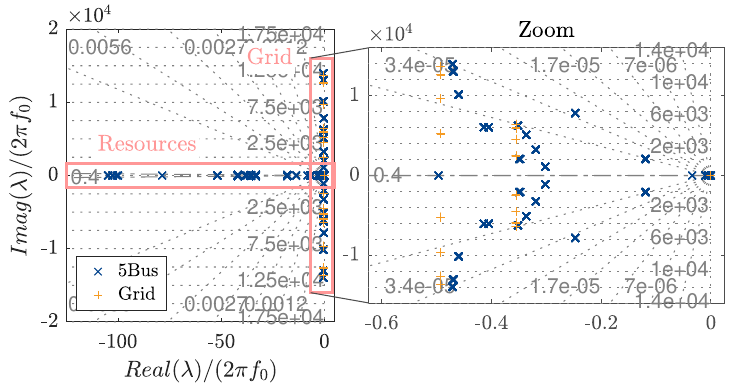}
	
	\caption
	{%
		Division of the system eigenvalues into the two sets associated with the grid and the resources, respectively.
		The left-hand side of the figure shows the entire region containing all eigenvalues, while the right-hand side figure shows a zoom on the grid eigenvalues.
		Additionally, the eigenvalues of the open-loop grid model are displayed.
	}
	\label{fig:hsa-sys}
\end{figure}

\cref{fig:hsa-sys} shows the comparison of the eigenvalues of the closed-loop system with those of the open-loop components.
To this end the eigenvalues are divided into two sets as marked by the two areas.

Additionally to the closed-loop eigenvalues of the system, \cref{fig:hsa-sys} displays the eigenvalues of the grid state-space model (i.e., the open-loop model).
One can see in the zoomed part of \cref{fig:hsa-sys} that the eigenvalues of the system (i.e., the closed-loop model) are shifted w.r.t.~those of the grid (i.e., the open-loop model).
Notably, most of these eigenvalues have fairly large imaginary parts (i.e, they occur at high frequencies).
These eigenvalues are related to the shunt capacitances of the lines, which are small (see \cref{tab:grid:parameters}).
By consequence, their shunt admittance becomes important only at high frequencies (i.e., since $Y=j\omega C$).

\begin{figure}[t]
	\centering
	\includegraphics[width=0.9\linewidth]{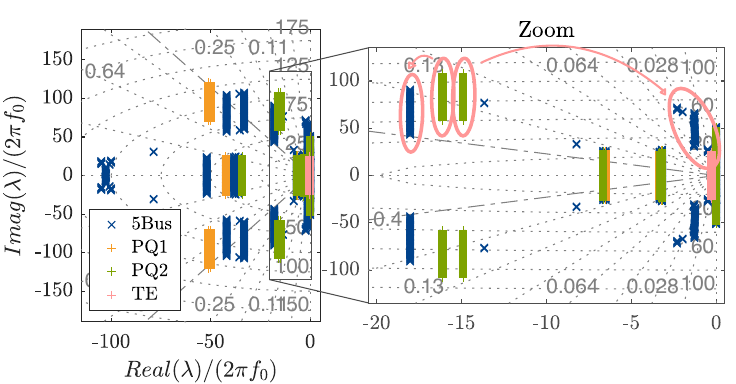}
	\caption
	{%
		Detailed view of the system eigenvalues associated with the resources.
		Additionally, the eigenvalues of the open-loop resource models are shown (i.e., PQ1/2 and \TE).
		Note the indication showing how this set of eigenvalues of PQ2 change their location for the system model.
	}
	\label{fig:hsa-sys:rsc-zoom:cmp}
\end{figure}

\cref{fig:hsa-sys:rsc-zoom:cmp} shows the portion of the system eigenvalues that are associated with the resources of the system.
For this reason, the eigenvalues of the two grid-following resources at node N04 and N05 and the \TE at node N01 are depicted too.
For the sake of clarity, all subsequent figures directly show the zoom on these eigenvalues.

As can be seen in \cref{fig:hsa-sys:rsc-zoom:cmp} the eigenvalues of the closed-loop system exhibit significant displacement as compared to those of the open-loop models.
In particular, the eigenvalues with low damping related to the \CIDER at N05 (see indications in \cref{fig:hsa-sys:rsc-zoom:cmp}) experience further reductions in the damping factor.

\begin{figure}[t]
	\centering
	\includegraphics[width=0.9\linewidth]{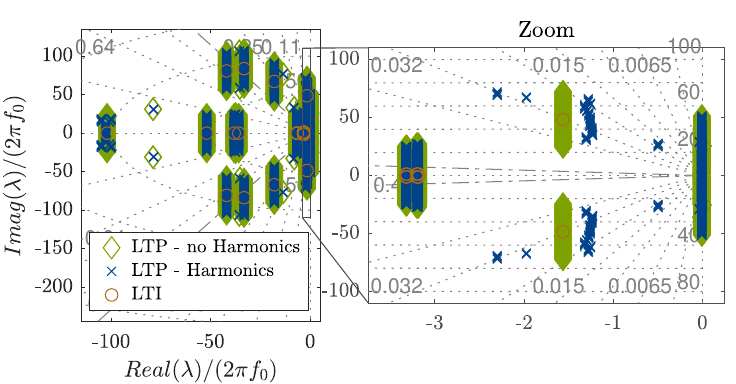}
	\caption
	{%
		Detailed view of the system eigenvalues associated with the resources.
		Comparison of the system eigenvalues obtained with the \HSS model considering a system with zero harmonic distortion to one with harmonic distortion.
		Additionally, the eigenvalues of the \LTI model are shown.
	}
	\label{fig:hsa-sys:rsc-zoom:h0-25}
\end{figure}

\cref{fig:hsa-sys:rsc-zoom:h0-25} illustrates the impact of harmonic distortion in a system on the location of the eigenvalues.
The eigenvalues of a system with harmonic distortion and a system with purely sinusoidal waveforms are compared.
The set of \LTP eigenvalues with low damping factor becomes more dispersed for the system with harmonic distortion.
This is due to the nonlinearity of the reference calculation of the \CIDER, which causes coupling between different frequencies (i.e., non-zero off-diagonal elements of the system matrix).
Since the reference calculation is not part of the internal response of the \CIDER[s], this effect can only be seen once the closed-loop system is calculated.
This is further underlined by the fact that the associated open-loop components in \cref{fig:hsa-sys:rsc-zoom:cmp} do not show this dispersion.

In \cref{fig:hsa-sys:rsc-zoom:h0-25}, the real parts of the eigenvalues of the \LTP system without harmonics match with those of the \LTI system.
This is according to expectations, since the \LTI and \LTP models of the \CIDER[s] are equivalent in the absence of harmonic distortion.
For the sake of clarity, the eigenvalues of the \LTI system are used subsequently to represent an \LTP system without harmonic distortion.

\subsection{Sensitivity Analysis and Harmonic Instability}
\label{sec:hsa-sys:instab}

\subsubsection{Methodology}

A sensitivity analysis of the system eigenvalues w.r.t.~to $K_{\ctrlFB,\act}$ of the \CIDER at N05 is performed.
The goal of this analysis is to find a case of harmonic instability.
Namely, an instability that occurs only if harmonic distortion is present in the system.
To this end, the controller gain is incrementally reduced by $1$\% of its previous value for $N =18$ iterations.
Recall from \cref{fig:hsa-rsc:sensitivity:flw}, that an increase of $K_{\ctrlFB,\act}$ yields better damping of the individual grid-following \CIDER.
Based on the eigenvalue loci, the stability boundary obtained from the \LTP and \LTI system eigenvalues are compared.

Finally, the same test is performed through \TDS with Simulink.
More precisely, two simulations are conducted in Simulink.
In both simulations, the feedback gain $K_{\ctrlFB,\act}$ of the \CIDER at N05 is decreased repeatedly over time.
The time between two changes is chosen long enough such that the system reaches a new steady state, before changing the parameter again.
The difference between the two simulations are the background harmonics injected at the substation.
In the first scenario, the \TE does not inject any harmonics, thus the approximate \LTI model is expected to yield accurate stability boundaries.
Conversely, in the second scenario, harmonics are injected as specified in \cref{tab:TE:harmonics}.
In this case, the more accurate \LTP model should outperform the \LTI model in terms of \HSA.

\subsubsection{Results and Discussion}

\begin{figure*}[t]
	\centering
		\includegraphics[width=0.9\linewidth]{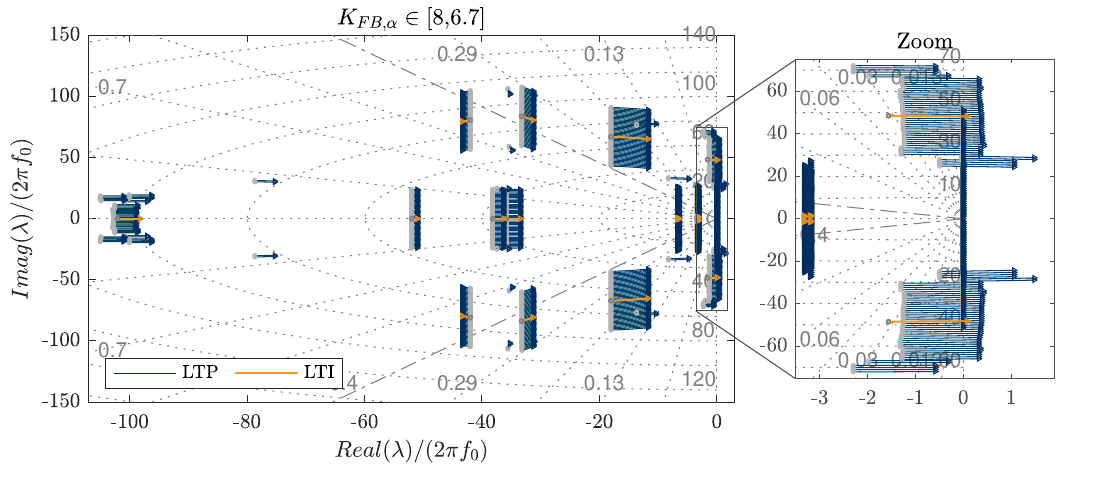}
	\caption
	{%
		Sensitivity analysis of the system eigenvalues for the \LTP and \LTI models w.r.t.~the controller gain $K_{\alpha,\ctrlFB }$ of the grid-following \CIDER at N05.
		The system eigenvalues associated with the resources are shown for a maximum harmonic order of $h_{max}=25$.
	}
	\label{fig:hsa-sys:sens}
\end{figure*}

\cref{fig:hsa-sys:sens} shows the sensitivity of the system eigenvalues w.r.t.~the controller gain $K_{\alpha,\ctrlFB }$ of the grid-following \CIDER at N05.
Note well that only the eigenvalues associated with the resources are affected by the tuning of the controller gain.
By contrast, the eigenvalues associated with the grid hardly move, since they are not part of the control loop.
For the sake of conciseness, this analysis is not shown in this paper, please see Chapter~5.5.4 of \cite{ths:2024:becker} for the details.

In \cref{fig:hsa-sys:sens} the eigenvalue loci, obtained through \LTP and \LTI models, are compared.
One can observe that the eigenvalues of the \LTP model cross the imaginary axis earlier than the ones of the \LTI model.
Thus, there exists a gain, which only leads to instability in case harmonic distortion is present in the system.
Furthermore, it is not possible to observe this instability with conventional stability criteria (i.e., the eigenvalues of the \LTI model).

\begin{figure}[t]
	\centering
	
	\subfloat[]
	{%
		\centering
		\includegraphics[width=0.9\linewidth]{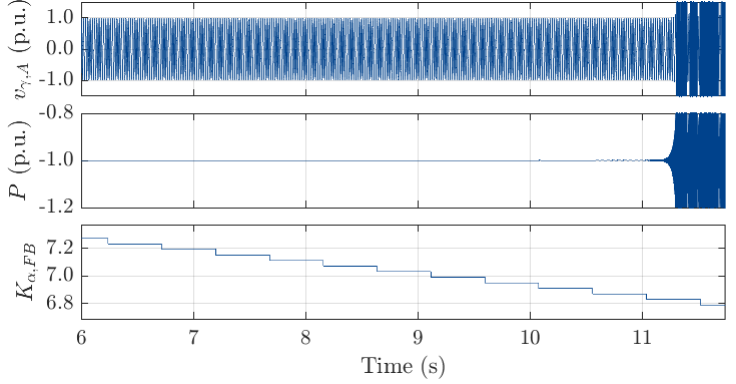}
		\label{fig:hsa-sys:tds:h1}
	}

	\subfloat[]
	{%
		\centering
		\includegraphics[width=0.9\linewidth]{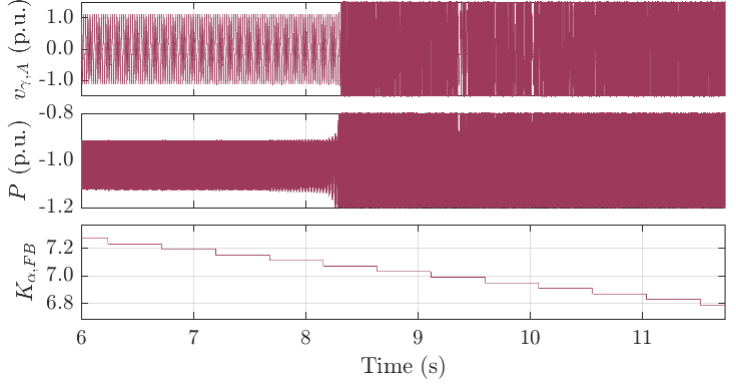}
		\label{fig:hsa-sys:tds:h25}
	}
	
	\caption
	{%
		Validation of the observed instability through \TDS in Simulink.
        \cref{fig:hsa-sys:tds:h1} shows stability boundary for the test system without harmonic injection at the substation (i.e., purely sinusoidal behaviour), and
		\cref{fig:hsa-sys:tds:h25} shows the system behaviour if harmonics are present (i.e., a ``harmonic instability'').
		Each figure depicts the nodal voltage of phase \phsA~(i.e., $v_{\grd,\phsA}$) and the injected power $P$ at node N05, in the first and second plot of the figure, respectively.
		The last plot in both figures indicates the decrease of the controller gain $K_{\alpha,\ctrlFB }$ of the grid-following \CIDER at N05 as the system becomes unstable.  
	}
	\label{fig:hsa-sys:tds}
\end{figure}

\cref{fig:hsa-sys:tds} shows the validation of the stability boundary observed in \cref{fig:hsa-sys:sens} through \TDS in Simulink.
In \cref{fig:hsa-sys:tds:h1}, the system is excited with a pure fundamental component at the substations, while in \cref{fig:hsa-sys:tds:h25}, the usual harmonic distortion is injected at the \TE.
Both subfigures show the nodal voltage and the injected active power at node N05 of the test system.
The last subplot in the subfigures shows how the controller gain $K_{\alpha,\ctrlFB }$ is decreased gradually until the system becomes unstable.
As can be seen in \cref{fig:hsa-sys:tds:h25}, the system becomes unstable earlier in the presence of harmonics, i.e., as compared to purely sinusoidal behaviour at the substation as illustrated \cref{fig:hsa-sys:tds:h1}.
This confirms the observations made in \cref{fig:hsa-sys:sens}, and shows a case of harmonic instability, that can not be observed with conventional stability criteria.
That is, an instability of this type can only be detected by the method which is based on \LTP models but not by one employing \LTI models.

\section{Conclusions}
\label{sec:conclusion}

In Part~II of this paper, the \HSA method proposed in Part~I was applied to both individual \CIDER[s] as well as entire microgrids.
The \HSS models of the individual \CIDER[s] were employed for detailed \HSA through eigenvalue analysis.
The impact of the maximum harmonic order on the eigenvalues, an in-depth analysis of their classification, and their sensitivity curves w.r.t.~to the \CIDER's control parameters was assessed.
The presented analyses allow to (i) understand the physical meaning and origin of the eigenvalues of a \CIDER (i.e., whether they are genuine to the model or spurious), and (ii) examine the stability characteristics of the \CIDER w.r.t. their control parameters and harmonic distortion.

Finally, the \HSA was demonstrated to accurately detect harmonic instability in a representative test system (i.e., with parameters based on the \CIGRE low-voltage benchmark microgrid~\cite{Rep:2014:CIGRE}).
The instability was identified through the system eigenvalues and confirmed with \TDS in Simulink.
Notably, the observed instability cannot be detected by the conventional \LTI modelling approach, which substantiates the value of the proposed framework for \HSA.
As mentioned in Part~I and presented in detail in previous papers of the authors \cite{jrn:2020:kettner-becker:HPF-1,jrn:2020:kettner-becker:HPF-2,jrn:2022:becker}, the modelling framework is both generic and modular.
That is, the proposed \HSA method can be employed for generic power systems with a large variety of \CIDER[s] and other typical power system components.

\appendices
\section{Reference Calculation of the Grid-Following CIDERs}
\label{app:cider:lib-rsc}

In grid-following \CIDER[s], the reference angle $\theta$ needed for the \cmpDQ~transform is provided by a synchronization unit, usually a \emph{Phase-Locked Loop} (\PLL).
The reference current $\IT^*_{\grd,\cmpDQ}$ is computed in order to track the power setpoint $S_\spt=P_\spt+jQ_\spt$ at the fundamental frequency.
In the vast majority of cases, synchronization units in general, and \PLL[s] in particular, are designed to lock to the fundamental positive-sequence component of the grid voltage (e.g., \cite{Bk:Teodorescu:2011}).
As a consequence, the grid current reference is calculated purely based on the $\cmpD$-component of the grid voltage:
\begin{equation}
	\IT^*_{\grd,\cmpDQ}(t)
	\approx  
	\begin{bmatrix}
		 \frac{1}{v_{\grd,\cmpD}(t)} & 0\\
		0 &  \frac{1}{v_{\grd,\cmpD}(t)}
	\end{bmatrix}
	\begin{bmatrix}
		P_\spt\\
		Q_\spt
	\end{bmatrix}
	\label{eq:ref:following:simplified}
\end{equation}

As required by power quality standards (e.g., \cite{Std:BSI-EN-50160:2000}), the grid voltages have to be maintained balanced and sinusoidal within specified limits.
Under these conditions, it reasonable to assume the following.
\begin{Hypothesis}\label{hyp:rsc:follower:harms}
	The time-variant signal content of $v_{\grd,\cmpD}(t)$, as given by $\xi_\cmpD(t)$ below, is low:
	\begin{alignat}{2}
		v_{\grd,\cmpD}(t)
		&=      V_{\grd,\cmpD,0}(1+\xi_\cmpD(t)),~
		&       \Abs{\xi_\cmpD(t)}
		&\ll    1
	\end{alignat}
    where $V_{\grd,\cmpD,0}$ is the Fourier coefficient of $v_{\grd,\cmpD}(t)$ at $h=0$.
\end{Hypothesis}

\subsection{Small-Signal Model of the Reference Calculation}
For the representation of the reference calculation within the \HSS model of the \CIDER, the format introduced in (12), Section~III.C of Part~I is employed.
To this end, one needs to derive the coefficient matrices $	\hat{\mathbf{R}}_{\refr}$ and $\hat{\mathbf{R}}_{\spt}$ originating from the Taylor expansion in time domain, as well as the shift of the origin $\hat{\bar{\mathbf{W}}}_{\ctrl}$.
Rewrite \eqref{eq:ref:following:simplified} in terms of the \CIDER variables:
\begin{align}
	\mathbf{w}_{\ctrl}(t)
	&=\frac{1}{\mathbf{w}_\refr(t)} \mathbf{w}_\spt(t)
\end{align}
Then, the coefficients for the small-signal model of \eqref{eq:ref:following:simplified} for the operating point $\mathbf{\bar{w}}_\refr(t)$ and a general power setpoint $\mathbf{\bar{w}}_\spt(t)$ are given by
\begin{align}
    \mathbf{\bar{w}}_{\ctrl}(t)
	&=\frac{1}{\mathbf{\bar{w}}_\refr(t)} \mathbf{\bar{w}}_\spt(t)
    \label{eq:refcalc:w_k}
    \\
	\mathbf{R}_{\refr}(t)
	&=-\left(\frac{1}{\mathbf{\bar{w}}_\refr(t)}\right)^2 \mathbf{\bar{w}}_\spt(t)
     \label{eq:refcalc:r_refr}
     \\
		\mathbf{R}_{\spt}(t)
	&=\frac{1}{\mathbf{\bar{w}}_\refr(t)}
    \label{eq:refcalc:r_spt}
\end{align}
From this, the harmonic-domain quantities $\hat{\mathbf{W}}_{\ctrl}$, $\hat{\mathbf{R}}_{\refr}$, and $\hat{\mathbf{R}}_{\spt}$ are derived by means of the Teoplitz theory.
To this end, the nonlinear elements of the above equations (i.e., the reciprocal of the grid voltage) need to be approximated in the harmonic domain.
%
%
\subsection{Nonlinear Approximation of the Reciprocal of the Grid Voltage}
Similar to the derivations of the reference calculation used in the \HPF in \cite{jrn:2020:kettner-becker:HPF-2}, the reciprocal of $\mathbf{\bar{w}}_\refr(t) = \bar{v}_{\grd,\cmpD}(t)$ can be approximated by a Taylor expansion.
In this respect, the following hypothesis is made:
\begin{Hypothesis}\label{hyp:rsc:follower:approx}
	For the calculation of the reference current in the grid-following \CIDER[s], the reciprocal of the grid voltage can be approximated by a  Taylor series of order $n$:
	\begin{equation}
		\frac{1}{\bar{v}_{\grd,\cmpD}(t)}
		\approx \frac{1}{\bar{V}_{\grd,\cmpD,0}}\sum_n(-1)^n\xi_\cmpD^{n}(t)
	\end{equation}
	where the Fourier series of $\xi_\cmpD(t)$ is described by
	\begin{equation}
	\xi_{\cmpD}(t) = \sum_{h\neq0}\frac{\bar{V}_{\grd,\cmpD,h}}{\bar{V}_{\grd,\cmpD,0}}\Exp{jh2\pi f_1 t}
	\end{equation}
\end{Hypothesis}
\noindent
Employing this Taylor expansion, one can approximate the matrix in \eqref{eq:ref:following:simplified} by $\boldsymbol{\Psi}^{(n)}(t)$
\begin{equation}
\boldsymbol{\Psi}^{(n)}(t)
= \frac{1}{\bar{V}_{\grd,\cmpD,0}}	\sum_n(-1)^n	\boldsymbol{\Xi}_\cmpD^{n}(t)
\label{eq:ref:td:nth-order}
\end{equation}
where
\begin{align}
	\boldsymbol{\Xi}_\cmpD(t) = \diag(\mathbf{1}_2)\xi_\cmpD(t)
\end{align}

Transforming \eqref{eq:ref:td:nth-order} to the harmonic domain gives
\begin{align}
\hat{\mathbf{\Psi}} ^{(n)}=  \frac{1}{\bar{V}_{\grd,\cmpD,0}}\sum_n(-1)^n\hat{\boldsymbol{\Xi}}_{\cmpD}^{n}
\label{eq:ref:hd:nth-order}
\end{align}
where $\diag(\mathbf{1})$ is a matrix of suitable size, composed of ones, and $\hat{\boldsymbol{\Xi}}_{\cmpD}$ describes the Toeplitz matrix built from the Fourier coefficients of $\mathbf{\Xi}_{\cmpD}(t)$.

Using a second-order Taylor expansion one obtains
\begin{align}
	\hat{\mathbf{\Psi}}^{(2)} =  \frac{1}{\bar{V}_{\grd,\cmpD,0}}\left(\diag(\mathbf{1})-\hat{\boldsymbol{\Xi}}_{\cmpD}+\hat{\boldsymbol{\Xi}}_{\cmpD}^2\right)
\end{align}
The impact of the order of the expansion on the accuracy of the \CIDER model was investigated in Chapter~3.4.2 of \cite{ths:2024:becker}.
It is shown that the second order expansion of the reference calculation provides good accuracy of the \HSS model as validated through \TDS.
Notably, expansions of higher order can easily be obtained using \eqref{eq:ref:hd:nth-order} with suitable $n$.
Note that the approximation of the reciprocal of the grid voltage is a nonlinear function of the Fourier coefficients of the grid voltage.




\bibliographystyle{IEEEtran}
\bibliography{Bibliography}

\end{document}